\documentclass[aps,prl,reprint,floatfix,superscriptaddress]{revtex4-1}
\usepackage{amsmath,amssymb}
\usepackage{graphicx}
\usepackage[utf8]{inputenc}
\usepackage{hyperref}
\usepackage{color}
\hypersetup{colorlinks=true,citecolor={blue},linkcolor={blue},urlcolor={blue}}
\usepackage{natbib}

\begin{document}

\title{Spontaneous symmetry breaking in a polariton and photon laser}

\author{H. Ohadi}
\author{E. Kammann}%
\affiliation{%
 School of Physics and Astronomy, University of Southampton, Southampton, SO17 1BJ, United Kingdom
}%
\author{T.C.H. Liew}
\affiliation{
Mediterranean Institute of Fundamental Physics, 31, via Appia Nuova, 00040, Rome, Italy
}%
\author{K.G. Lagoudakis}
\affiliation{
Stanford University, 348 Via Pueblo Mall, Stanford, CA 94305-4088, US
}
\author{A.V. Kavokin}%
\affiliation{
Spin Optics Laboratory, St. Petersburg State University, 1, Ulianovskaya, St. Petersburg, 198504, Russia
}
\affiliation{%
 School of Physics and Astronomy, University of Southampton, Southampton, SO17 1BJ, United Kingdom
}%
\author{P.G. Lagoudakis}
\email[correspondence address: ]{pavlos.lagoudakis@soton.ac.uk}
\affiliation{%
 School of Physics and Astronomy, University of Southampton, Southampton, SO17 1BJ, United Kingdom
}%

\date{\today}

\begin{abstract} We report on the simultaneous observation of spontaneous
symmetry breaking and long-range spatial coherence both in the strong and the
weak-coupling regime in a semiconductor microcavity. Under pulsed excitation,
the formation of a stochastic order parameter is observed in polariton
and photon lasing regimes. Single-shot measurements of the Stokes vector of the
emission exhibit the buildup of stochastic polarization. Below threshold, the
polarization noise does not exceed 10\%, while above threshold we observe a
total polarization of up to 50\% after each excitation pulse, while the
polarization averaged over the ensemble of pulses remains nearly zero. In both
polariton and photon lasing regimes, the stochastic polarization buildup is
accompanied by the buildup of spatial coherence. We find that the Landau
criterion of spontaneous symmetry breaking and Penrose-Onsager criterion of
long-range order for Bose-Einstein condensation are met in both polariton and
photon lasing regimes.

\end{abstract}


\maketitle

Bosonic particles cooled below the temperature of quantum degeneracy undergo a
thermodynamic phase transition during which phase correlations spontaneously
build up and the system enters a macroscopic coherent state. This process is
known as Bose-Einstein condensation (BEC). BEC is characterised by a buildup of
an order parameter, which is usually associated with a macroscopic
multi-particle wave function of the
condensate~\cite{penrose_bose-einstein_1956}. Recently, observation of BEC of
exciton-polaritons in CdTe, GaAs and GaN-based microcavities has been claimed by
several groups~\cite{kasprzak_bose-einstein_2006, balili_bose-einstein_2007,
deng_condensation_2002, baumberg_spontaneous_2008}.  Several features of BEC
have indeed been observed, such as: the macroscopic occupation of a quantum
state on top of a Boltzmann-like distribution, the long-range spatial coherence,
the spectral narrowing of the photoluminescence signal and the decrease of the
second order coherence. While necessary, these signatures are not unambiguous
proofs of BEC in exciton-polariton planar microcavities simply due to the fact
that these features have also been reported in conventional lasers [vertical
cavity surface emitting lasers 
(VCSELs)]~\cite{van_exter_effect_1995,jin_physics_vcsels}. The terminology used in
describing degenerate polariton gases is still a subject of debate in the
scientific community ranging from polariton BEC to polariton
laser~\cite{butov_behaviour_2012,deveaud-pledran_condensation_2012}. More
recently BEC of photons was reported in a weakly-coupled dye-filled
microcavity~\cite{klaers_boseeinstein_2010}. However, unlike in strongly-coupled
microcavities~\cite{baumberg_spontaneous_2008}, spontaneous symmetry breaking
which is a characteristic feature of the BEC phase transition has not been
previously observed in weakly-coupled microcavities. 
 
Here we report on the spontaneous buildup of stochastic polarization vector and
long-range coherence in a GaAs quantum-well microcavity, in the strong-coupling
regime, where we observe polariton lasing and, surprisingly, also in the
weak-coupling photon lasing regime. Previous studies in this sample have shown
both photon lasing under continuous wave excitation~\cite{bajoni_photon_2007}
and polariton lasing under pulsed excitation~\cite{maragkou_longitudinal_2010}.
Under high-excitation densities, we have recently observed a dynamical crossover
between photon and polariton lasing following a single excitation pulse~\cite{
kammann_crossover_2011}. This is in agreement with previous observations where
increasing carrier density was shown to transfer the system from strong to weak
coupling~\cite{nelsen_lasing_2009,tempel_characterization_2012}.

\begin{figure*}
\includegraphics[width=0.9\textwidth]{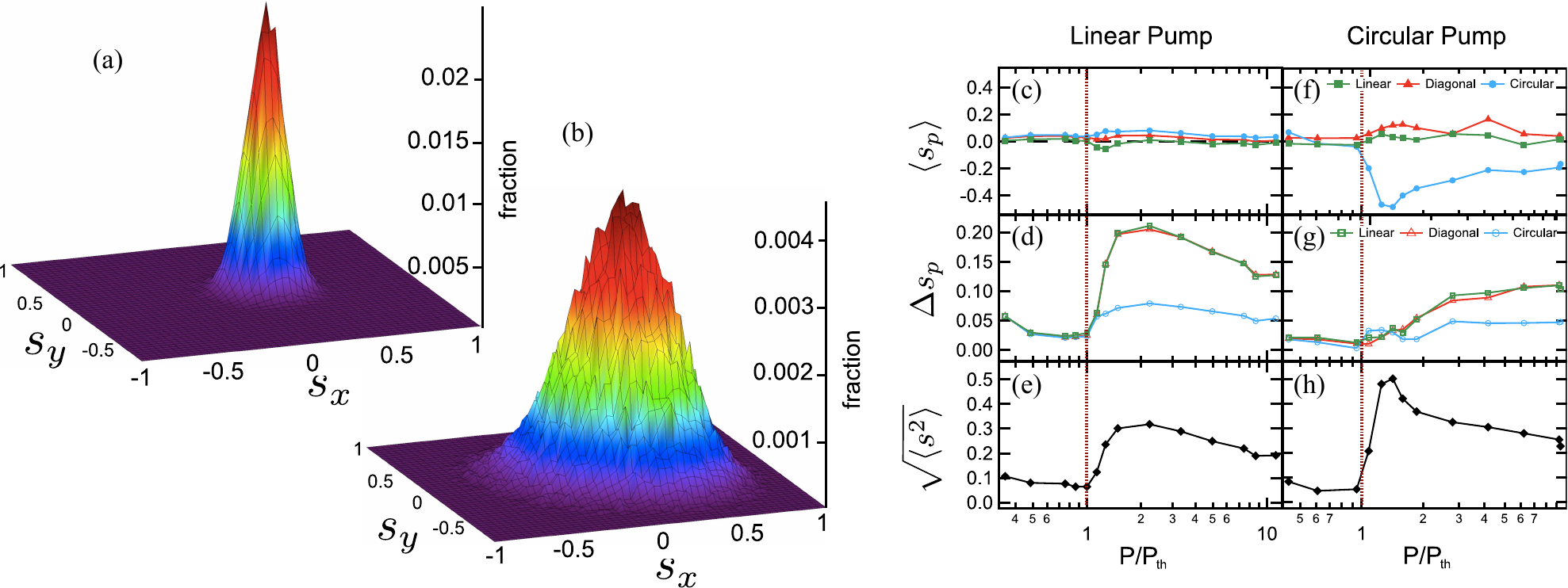}%
\caption{\label{fig:hills_total_degree}2D histogram of linear and diagonal components of the
pseudo-spin vector for $P < P_{th}$ (a) and $P = 2.4 P_{th}$ (b) on the $x-y$ plane of the
Stokes sphere. The mean (c,~f), the standard deviation (d,~g) and the total
degree of polarization (e,~h) for linearly (left panel) and circularly (right
panel) polarized pump. Above threshold (marked by a red line) we observe the
buildup of total degree of polarization while the average polarization remains
close to zero for linearly polarized pump.}
\end{figure*}

The total angular momentum $\mathbf{J}$ of a heavy-hole exciton-polariton in a
quantum-well microcavity can have two possible  projections on the growth axis
of the structure, $J_z=+1$  (spin up) or $J_z=-1$  (spin down), corresponding to
right and left circularly polarized emitted light, respectively. The order
parameter for an exciton-polariton condensate can then be defined by a
two-component complex vector
$\psi_\sigma(\mathbf{r},t)=[\psi_{+1}(\mathbf{r},t),\psi_{-1}(\mathbf{r},t)]$,
where $\psi_{+1}(\mathbf{r},t)$ and $\psi_{-1}(\mathbf{r},t)$ are multi-particle
wave functions of the spin-up and spin-down components of the condensate. The
components of the complex order parameter $\psi_\sigma$ are related to the
measurable condensate pseudo-spin (Stokes vector) $\mathbf S$ by
$S_x=\mathrm{Re}(\psi_{-1}^*\psi_{+1})$,
$S_y=\mathrm{Im}(\psi_{-1}^*\psi_{+1})$,
$S_z=(\vert\psi_{+1}\vert^2-\vert\psi_{-1}\vert^2)/2$. The components of the
Stokes vector are proportional to the linear, diagonal and circular polarization
degree of light emitted by the condensate.  These polarization degrees define
the normalized pseudo-spin vector $\mathbf s=2\overline{\mathbf{S}}/n$, where
$n$ is the occupation number of the condensate. This vector can be quantified
experimentally by polarized photoluminescence (PL)
measurements~\cite{lagoudakis_stimulated_2002,kavokin_polarization_2003}.  It
contains important information on the order parameter of the condensate. The
total degree of polarization of the emitted light given by the amplitude of
$\mathbf s$ changes from zero for a chaotic state to 1 for an ideal condensate.
The spontaneous symmetry breaking of the condensate manifests itself in the
spontaneous buildup of polarization of the emitted
light~\cite{laussy_effects_2006}. In the absence of pinning, this polarization
is randomly chosen by the system and changes stochastically from one experiment
to another~\cite{read_stochastic_2009}.

The details for the microcavity structure could be found in
Ref.~\cite{bajoni_photon_2007} and the experimental setup is explained in detail
in the Supplemental Material. About $64,000$ shots are recorded and analysed
for different polarizations and powers of the excitation laser. The components
of the normalized pseudo-spin vector $\mathbf s$ are measured by
$s_{x,y,z}=(I_{0^\circ,45^\circ,\circlearrowleft}-I_{90^\circ,-45^\circ,\circlearrowright})/(I_{0^\circ,45^\circ,\circlearrowleft}+I_{90^\circ,-45^\circ,\circlearrowright})$,
where $I$ is the peak intensity of the PL measured to an accuracy of $> 94 \%$.
polarization distributions were deconvoluted with the intrinsic noise of the
detection system. The PL can be spectrally filtered by means of a grating and an
adjustable slit before focusing into the detectors.

Here we measure the polarization of emitted light after each
excitation pulse. This measurement is integrated over the lifetime of each
individual condensate excited by a laser pulse. On the other hand, we do not
average on the ensemble of pulses, but we perform single-shot polarization
measurements. Figure~\ref{fig:hills_total_degree}(a,b) shows the 2D projection of
the Stokes parameter histogram on the $x-y$ plane (linear polarization plane)
for $\sim 64,000$ shots. In the linear regime and when the pump power is below
threshold [Fig.~\ref{fig:hills_total_degree}(a)] we observe a narrow Gaussian
distribution centred to zero, which demonstrates the unpolarized nature of the
polariton PL. The width of the Gaussian distribution is limited by the analogue
noise and the finite digital sampling rate of the acquisition. Above threshold
($P = 2.4 P_{th}$) and still in the strong-coupling regime
[Fig.~\ref{fig:hills_total_degree}(b)] the distribution significantly broadens,
with the average polarization again remaining close to zero. Although the
average polarization in the strong-coupling regime is zero we see that each
realisation of the condensate is indeed polarized with a random orientation of
the Stokes vector.

Fig.~\ref{fig:hills_total_degree}(c-h) shows the mean degree of polarization, the
standard deviation ($\Delta s$) and average degree of polarization for linear
and circular pump polarization.  The increase of standard deviation of
polarization in the ensemble of shots is a manifestation of the stochastic
buildup of polarization. Contrary to other microcavity
samples~\cite{kasprzak_build_2007,christmann_room_2008} we do not observe any
polarization pinning to the crystallographic axes and below threshold the
polarization is unpolarized. The total degree of polarization increases above
threshold followed by a sharp drop with increasing pump power.

In the strong-coupling regime the density of electron-hole pairs remains below
the Mott transition, where the Coulomb attraction is sufficient to stabilise
Wannier-Mott excitons. By increasing the excitation beyond the Mott transition
density the excitons ionize and an electron-hole plasma is
formed~\cite{kappei_direct_2005,stern_mott_2008}. Energy-resolved Fourier
space imaging allows for direct evidence of both the strong and the
weak-coupling regime as the corresponding dispersion relations ($E(k)$) are
fundamentally different. Above threshold and still in the strong-coupling regime
($P = 1.1 P_{th}$) we observe a blue-shifted lower exciton-polariton emission and as
we increase the power ($P > 4P_{th}$) we enter the weak-coupling regime and an
emission peak near the bare cavity photon energy
emerges~\cite{kammann_crossover_2011}.

\begin{figure} \includegraphics[width=0.45\textwidth]{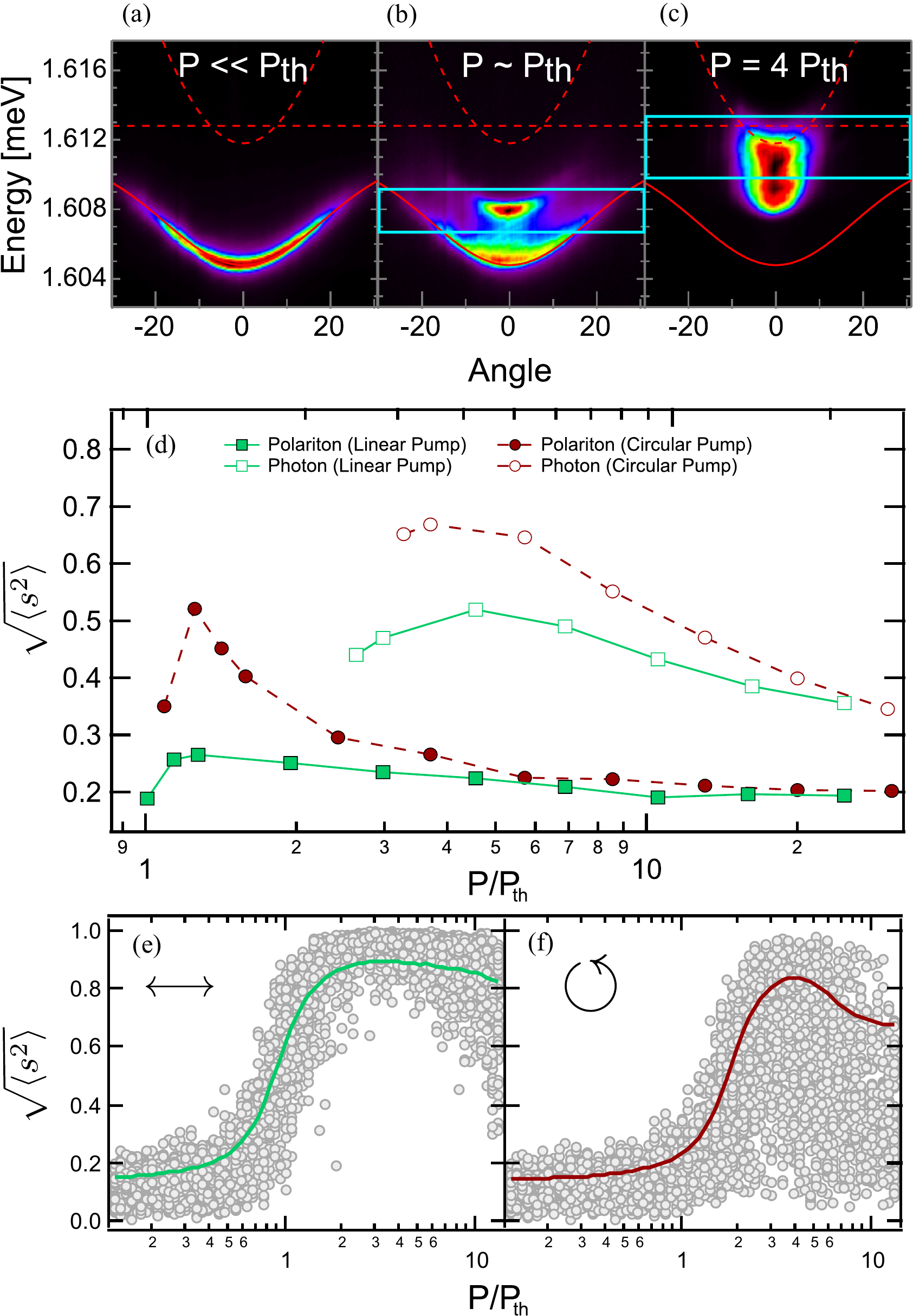}%
\caption{\label{fig:dispersion_e_resolv}Dispersion images for (a) below
threshold (linear regime), (b) $P = 1.5 P_{th}$ (strong-coupling regime) and (c)
$P = 4 P_{th}$ (weak-coupling regime). The cyan border marks the regions that
are filtered for polariton states (b) and bare cavity photon states (c). (d) The
total degree of polarization for circularly polarized (dotted lines) and
linearly polarized pump (solid lines) for photon (open circles) and polariton
energy states (closed circles). (e), (f) Simulation results for the dependence
of the total polarization degree on pump intensity for a linearly polarized (e)
and circularly polarized (f) pump. The grey circles show the values for each
pulse and the solid curves show the root mean square.} \end{figure}

Using a grating and a variable mechanical slit we can spectrally filter the
photoluminescence emission of polaritons (strong coupling) or bare cavity photon
energy states (weak coupling). Fig.~\ref{fig:dispersion_e_resolv}(a-c) shows
the dispersion for the linear, strong-coupling and weak-coupling regime
respectively. The cyan border displays the energy gap that is filtered for the
measurement.  Fig.~\ref{fig:dispersion_e_resolv}(d) shows the total degree of
polarization of photon and polariton energy states for linearly and circularly
polarized pump. The observed total degree of polarization at bare cavity photon
energy is twice higher than that of the polariton states.

\begin{figure*}
\includegraphics[width=1\textwidth]{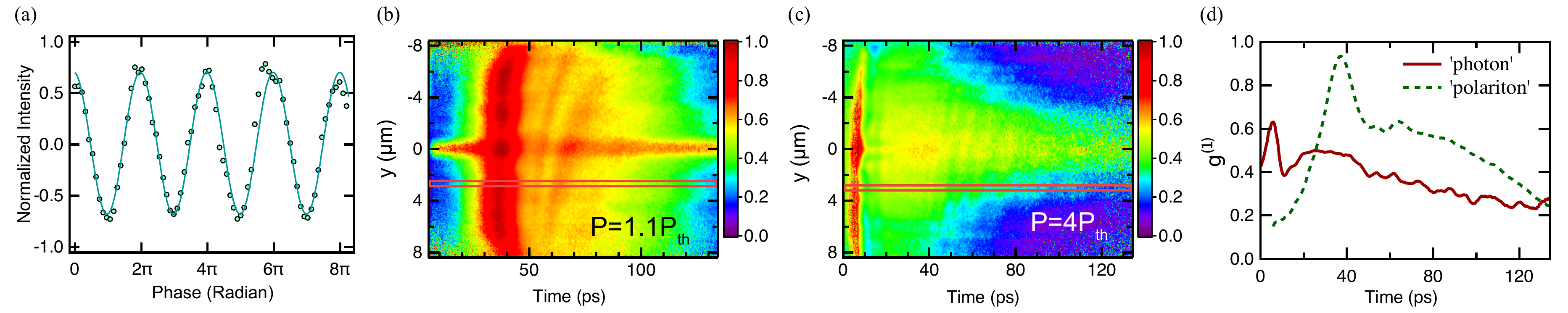}
\caption{
\label{fig:TR}
(a) Normalized intensity on one pixel as a function of phase between the arms of
the interferometer. The first-order coherence $g^{(1)}(\mathbf r, \mathbf{-r})$
is given by the amplitude. Time resolved first-order spatial coherence, taken
across the autocorrelation point at (b) $P=1.1 P_{th}$ and (c) $P=4 P_{th}$. (d)
The profile of the first-order coherence as a function of time in polariton and photon laser regimes, at regions marked in red in (b) and (c).
}
\end{figure*}

A theoretical description of the stochastic polarization buildup in the strong
coupling regime can be based on the development of the Langevin equation for
exciton-polaritons~\cite{read_stochastic_2009}, which replaces the Fokker-Planck
equation for distribution functions with a stochastic element. Within this
theory, which is described in detail in the Supplementary Information, each
pulse in the experiment is modelled separately and receives an influx of
polaritons with random polarization from the hot incoherent excitons excited by
the nonresonant pump.  Below threshold, the polariton polarization changes
rapidly within the duration of each pulse since polaritons excited in the system
have uncorrelated stochastic polarization. Equivalently, there is no coherence
between polaritons and the total polarization degree (whether averaged over
multiple pulses or not) is minimal.

Above threshold, stimulated scattering amplifies the polarization state of
polaritons. Polaritons become coherent and a high total polarization degree is
expected in each pulse. In the absence of pinning, this polarization is randomly
chosen by the system and changes stochastically from one pulse to another. This
effect appears both experimentally [Fig.~\ref{fig:dispersion_e_resolv}(d)] and
theoretically [see Fig.~\ref{fig:dispersion_e_resolv}(f)], however, our theory
predicts larger polarization degrees than those observed experimentally. This is
to be expected since the theory does not take into account the spatial
inhomogeneity of the condensate.

It is important to note that since the average linear polarization degrees
observed in our system are minimal [Fig.~\ref{fig:hills_total_degree}(c)], we
expect that only a small polarization splitting is present in the sample. For
this reason there is minimal spin relaxation or rotation of the Stokes vector,
at moderate densities, after the polarized polariton condensate has formed in
each pulse. However, for high polariton densities the anisotropy of
polariton-polariton interactions introduces a nonlinear mechanism of Stokes
vector precession known as self-induced Larmor
precession~\cite{shelykh_semiconductor_2004}. This mechanism causes the
reduction of the total polarization degree, which is time integrated over the
duration of each pulse, for increasing pump intensity in both experiment and
theory.

Under circularly polarized pumping the presence of a nonzero circular
polarization degree of the polariton condensate demonstrates incomplete spin
relaxation of the excited carriers in our experiment
[Fig.~\ref{fig:hills_total_degree}(f)]. Theoretically, this is
accounted for by a partial spin relaxation between hot exciton reservoirs. Note
that the incomplete spin relaxation does not imply that any phase coherence is
preserved from the nonresonant laser.

Our theoretical results demonstrate a lower polarization degree with circularly
polarized pumping than with linearly polarized pumping. This is due to increased
self-induced Larmor precession when there is a larger imbalance in the spin
populations. However, experimentally we rather observe that the polarization
degree is lower with linearly polarized pumping. This difference can also be
attributed to the spatial degree of freedom, which is not accounted for in our
theoretical model. Under linearly polarized pumping, different points in the
sample may be polarized differently, in particular since they can have different
angles of linear polarization. Under circularly polarized pump, we expect
smaller spatial variations in polarization, since the circular polarization is
preferred at all points in the sample.

To reveal the buildup of coherence, the first-order spatial correlation
function ($g^{(1)}(\mathbf{r}, \mathbf{-r})$) of the polariton emission is
studied using an actively-stabilised Michelson interferometer in a
mirror-retro-reflector configuration with close to zero arm length difference (for details see
Ref.~\cite{kasprzak_bose-einstein_2006}). To time-resolve the correlation
function a diffraction-limited 1-dimensional stripe of the interferograms is
sent to a streak camera~\cite{nardin_dynamics_2009}. By scanning the phase
between the mirror and the retro-reflector arm [Fig.~\ref{fig:TR}(a)]
$g^{(1)}(\mathbf r, \mathbf{-r})$ can be extracted. Time-resolved real-space
correlation functions are shown in Figs.~\ref{fig:TR}(b) and Fig.~\ref{fig:TR}(c) for
$P = 1.5 P_{th}$ and $P = 4 P_{th}$ respectively.  Fig.~\ref{fig:TR}(b) shows the
macroscopic phase coherence for the exciton-polariton condensate. In
Fig.~\ref{fig:TR}(c) we observe a high degree of coherence for photon lasing along
the whole real space profile. In the transition to polariton lasing the
coherence drops significantly and builds up again when the system enters the
strong-coupling [Fig.~\ref{fig:TR}(d)], but with a lower degree of coherence due
to the lower occupation of the ground state~\cite{kammann_crossover_2011}. 

In Ref.~\cite{baumberg_spontaneous_2008} GaN bulk microcavity which is a
strongly spatially inhomogeneous system was studied. While this system did show
polariton lasing at room temperature and spontaneous vector polarization, it did
not show a long-range spatial order, which according to the Penrose-Onsager
criterion~\cite{penrose_bose-einstein_1956} is a characteristic feature of the
phase transition toward Bose-Einstein condensation. The long-range spatial
order in polariton condensates was demonstrated by Kasprzak \emph{et~al}.
\cite{kasprzak_bose-einstein_2006}. However, they did not observe a stochastic
polarization, another key feature of the phase transition which is the
spontaneous symmetry breaking. To our knowledge, our present work is the first
to show the spontaneous stochastic polarization and long-range spatial coherence
in the same structure. We demonstrate that these two crucial features of the
phase transition can be observed both in the strong and weak-coupling regime and
show theoretical agreement for the exciton-polariton regime.

\begin{acknowledgments} P.G.L. and A.V.K. acknowledge the FP7 ITN-Spinoptronics,
  ITN-Clermont 4, Royal Society and EPSRC through contract EP/F026455/1 for
  funding. T.C.H.L. acknowledges the support of the European Commission under
  the Marie Curie Intra-European Fellowship EPOQUES.  The authors thank
  Jacqueline Bloch and Aristide Lema\^{i}tre for the provision of the sample.
\end{acknowledgments}

\bibliography{main}

\begin{thebibliography}{25}%
\makeatletter
\providecommand \@ifxundefined [1]{%
 \@ifx{#1\undefined}
}%
\providecommand \@ifnum [1]{%
 \ifnum #1\expandafter \@firstoftwo
 \else \expandafter \@secondoftwo
 \fi
}%
\providecommand \@ifx [1]{%
 \ifx #1\expandafter \@firstoftwo
 \else \expandafter \@secondoftwo
 \fi
}%
\providecommand \natexlab [1]{#1}%
\providecommand \enquote  [1]{``#1''}%
\providecommand \bibnamefont  [1]{#1}%
\providecommand \bibfnamefont [1]{#1}%
\providecommand \citenamefont [1]{#1}%
\providecommand \href@noop [0]{\@secondoftwo}%
\providecommand \href [0]{\begingroup \@sanitize@url \@href}%
\providecommand \@href[1]{\@@startlink{#1}\@@href}%
\providecommand \@@href[1]{\endgroup#1\@@endlink}%
\providecommand \@sanitize@url [0]{\catcode `\\12\catcode `\$12\catcode
  `\&12\catcode `\#12\catcode `\^12\catcode `\_12\catcode `\%12\relax}%
\providecommand \@@startlink[1]{}%
\providecommand \@@endlink[0]{}%
\providecommand \url  [0]{\begingroup\@sanitize@url \@url }%
\providecommand \@url [1]{\endgroup\@href {#1}{\urlprefix }}%
\providecommand \urlprefix  [0]{URL }%
\providecommand \Eprint [0]{\href }%
\providecommand \doibase [0]{http://dx.doi.org/}%
\providecommand \selectlanguage [0]{\@gobble}%
\providecommand \bibinfo  [0]{\@secondoftwo}%
\providecommand \bibfield  [0]{\@secondoftwo}%
\providecommand \translation [1]{[#1]}%
\providecommand \BibitemOpen [0]{}%
\providecommand \bibitemStop [0]{}%
\providecommand \bibitemNoStop [0]{.\EOS\space}%
\providecommand \EOS [0]{\spacefactor3000\relax}%
\providecommand \BibitemShut  [1]{\csname bibitem#1\endcsname}%
\let\auto@bib@innerbib\@empty
\bibitem [{\citenamefont {Penrose}\ and\ \citenamefont
  {Onsager}(1956)}]{penrose_bose-einstein_1956}%
  \BibitemOpen
  \bibfield  {author} {\bibinfo {author} {\bibfnamefont {O.}~\bibnamefont
  {Penrose}}\ and\ \bibinfo {author} {\bibfnamefont {L.}~\bibnamefont
  {Onsager}},\ }\href {\doibase 10.1103/PhysRev.104.576} {\bibfield  {journal}
  {\bibinfo  {journal} {Phys. Rev.}\ }\textbf {\bibinfo {volume} {104}},\
  \bibinfo {pages} {576} (\bibinfo {year} {1956})}\BibitemShut {NoStop}%
\bibitem [{\citenamefont {Kasprzak}\ \emph {et~al.}(2006)\citenamefont
  {Kasprzak}, \citenamefont {Richard}, \citenamefont {Kundermann},
  \citenamefont {Baas}, \citenamefont {Jeambrun}, \citenamefont {Keeling},
  \citenamefont {Marchetti}, \citenamefont {Szymańska}, \citenamefont
  {André}, \citenamefont {Staehli}, \citenamefont {Savona}, \citenamefont
  {Littlewood}, \citenamefont {Deveaud},\ and\ \citenamefont
  {Dang}}]{kasprzak_bose-einstein_2006}%
  \BibitemOpen
  \bibfield  {author} {\bibinfo {author} {\bibfnamefont {J.}~\bibnamefont
  {Kasprzak}}, \bibinfo {author} {\bibfnamefont {M.}~\bibnamefont {Richard}},
  \bibinfo {author} {\bibfnamefont {S.}~\bibnamefont {Kundermann}}, \bibinfo
  {author} {\bibfnamefont {A.}~\bibnamefont {Baas}}, \bibinfo {author}
  {\bibfnamefont {P.}~\bibnamefont {Jeambrun}}, \bibinfo {author}
  {\bibfnamefont {J.~M.~J.}\ \bibnamefont {Keeling}}, \bibinfo {author}
  {\bibfnamefont {F.~M.}\ \bibnamefont {Marchetti}}, \bibinfo {author}
  {\bibfnamefont {M.~H.}\ \bibnamefont {Szymańska}}, \bibinfo {author}
  {\bibfnamefont {R.}~\bibnamefont {André}}, \bibinfo {author} {\bibfnamefont
  {J.~L.}\ \bibnamefont {Staehli}}, \bibinfo {author} {\bibfnamefont
  {V.}~\bibnamefont {Savona}}, \bibinfo {author} {\bibfnamefont {P.~B.}\
  \bibnamefont {Littlewood}}, \bibinfo {author} {\bibfnamefont
  {B.}~\bibnamefont {Deveaud}}, \ and\ \bibinfo {author} {\bibfnamefont
  {L.~S.}\ \bibnamefont {Dang}},\ }\href {\doibase 10.1038/nature05131}
  {\bibfield  {journal} {\bibinfo  {journal} {Nature}\ }\textbf {\bibinfo
  {volume} {443}},\ \bibinfo {pages} {409–14} (\bibinfo {year}
  {2006})}\BibitemShut {NoStop}%
\bibitem [{\citenamefont {Balili}\ \emph {et~al.}(2007)\citenamefont {Balili},
  \citenamefont {Hartwell}, \citenamefont {Snoke}, \citenamefont {Pfeiffer},\
  and\ \citenamefont {West}}]{balili_bose-einstein_2007}%
  \BibitemOpen
  \bibfield  {author} {\bibinfo {author} {\bibfnamefont {R.}~\bibnamefont
  {Balili}}, \bibinfo {author} {\bibfnamefont {V.}~\bibnamefont {Hartwell}},
  \bibinfo {author} {\bibfnamefont {D.}~\bibnamefont {Snoke}}, \bibinfo
  {author} {\bibfnamefont {L.}~\bibnamefont {Pfeiffer}}, \ and\ \bibinfo
  {author} {\bibfnamefont {K.}~\bibnamefont {West}},\ }\href {\doibase
  10.1126/science.1140990} {\bibfield  {journal} {\bibinfo  {journal}
  {Science}\ }\textbf {\bibinfo {volume} {316}},\ \bibinfo {pages} {1007 }
  (\bibinfo {year} {2007})}\BibitemShut {NoStop}%
\bibitem [{\citenamefont {Deng}\ \emph {et~al.}(2002)\citenamefont {Deng},
  \citenamefont {Weihs}, \citenamefont {Santori}, \citenamefont {Bloch},\ and\
  \citenamefont {Yamamoto}}]{deng_condensation_2002}%
  \BibitemOpen
  \bibfield  {author} {\bibinfo {author} {\bibfnamefont {H.}~\bibnamefont
  {Deng}}, \bibinfo {author} {\bibfnamefont {G.}~\bibnamefont {Weihs}},
  \bibinfo {author} {\bibfnamefont {C.}~\bibnamefont {Santori}}, \bibinfo
  {author} {\bibfnamefont {J.}~\bibnamefont {Bloch}}, \ and\ \bibinfo {author}
  {\bibfnamefont {Y.}~\bibnamefont {Yamamoto}},\ }\href {\doibase
  10.1126/science.1074464} {\bibfield  {journal} {\bibinfo  {journal}
  {Science}\ }\textbf {\bibinfo {volume} {298}},\ \bibinfo {pages} {199 }
  (\bibinfo {year} {2002})}\BibitemShut {NoStop}%
\bibitem [{\citenamefont {Baumberg}\ \emph {et~al.}(2008)\citenamefont
  {Baumberg}, \citenamefont {Kavokin}, \citenamefont {Christopoulos},
  \citenamefont {Grundy}, \citenamefont {Butté}, \citenamefont {Christmann},
  \citenamefont {Solnyshkov}, \citenamefont {Malpuech}, \citenamefont
  {Baldassarri Höger~von Högersthal}, \citenamefont {Feltin}, \citenamefont
  {Carlin},\ and\ \citenamefont {Grandjean}}]{baumberg_spontaneous_2008}%
  \BibitemOpen
  \bibfield  {author} {\bibinfo {author} {\bibfnamefont {J.~J.}\ \bibnamefont
  {Baumberg}}, \bibinfo {author} {\bibfnamefont {A.~V.}\ \bibnamefont
  {Kavokin}}, \bibinfo {author} {\bibfnamefont {S.}~\bibnamefont
  {Christopoulos}}, \bibinfo {author} {\bibfnamefont {A.~J.~D.}\ \bibnamefont
  {Grundy}}, \bibinfo {author} {\bibfnamefont {R.}~\bibnamefont {Butté}},
  \bibinfo {author} {\bibfnamefont {G.}~\bibnamefont {Christmann}}, \bibinfo
  {author} {\bibfnamefont {D.~D.}\ \bibnamefont {Solnyshkov}}, \bibinfo
  {author} {\bibfnamefont {G.}~\bibnamefont {Malpuech}}, \bibinfo {author}
  {\bibfnamefont {G.}~\bibnamefont {Baldassarri Höger~von Högersthal}},
  \bibinfo {author} {\bibfnamefont {E.}~\bibnamefont {Feltin}}, \bibinfo
  {author} {\bibfnamefont {J.}~\bibnamefont {Carlin}}, \ and\ \bibinfo {author}
  {\bibfnamefont {N.}~\bibnamefont {Grandjean}},\ }\href {\doibase
  10.1103/PhysRevLett.101.136409} {\bibfield  {journal} {\bibinfo  {journal}
  {Phys.~Rev.~Lett.}\ }\textbf {\bibinfo {volume} {101}},\ \bibinfo {pages}
  {136409} (\bibinfo {year} {2008})}\BibitemShut {NoStop}%
\bibitem [{\citenamefont {van Exter}\ \emph {et~al.}(1995)\citenamefont {van
  Exter}, \citenamefont {Jansen Van~Doorn},\ and\ \citenamefont
  {Woerdman}}]{van_exter_effect_1995}%
  \BibitemOpen
  \bibfield  {author} {\bibinfo {author} {\bibfnamefont {M.}~\bibnamefont {van
  Exter}}, \bibinfo {author} {\bibfnamefont {A.}~\bibnamefont {Jansen
  Van~Doorn}}, \ and\ \bibinfo {author} {\bibfnamefont {J.}~\bibnamefont
  {Woerdman}},\ }\href {\doibase 10.1109/2944.401247} {\bibfield  {journal}
  {\bibinfo  {journal} {IEEE J. Sel. Top. Quantum Electron.}\ }\textbf
  {\bibinfo {volume} {1}},\ \bibinfo {pages} {601 } (\bibinfo {year}
  {1995})}\BibitemShut {NoStop}%
\bibitem [{\citenamefont {Jin}\ \emph {et~al.}(1995)\citenamefont {Jin},
  \citenamefont {Khitrova}, \citenamefont {Boggavarapu}, \citenamefont {Gibbs},
  \citenamefont {Koch},\ and\ \citenamefont {Tobin}}]{jin_physics_vcsels}%
  \BibitemOpen
  \bibfield  {author} {\bibinfo {author} {\bibfnamefont {R.}~\bibnamefont
  {Jin}}, \bibinfo {author} {\bibfnamefont {G.}~\bibnamefont {Khitrova}},
  \bibinfo {author} {\bibfnamefont {D.}~\bibnamefont {Boggavarapu}}, \bibinfo
  {author} {\bibfnamefont {H.~M.}\ \bibnamefont {Gibbs}}, \bibinfo {author}
  {\bibfnamefont {S.~W.}\ \bibnamefont {Koch}}, \ and\ \bibinfo {author}
  {\bibfnamefont {M.~S.}\ \bibnamefont {Tobin}},\ }\href {\doibase
  10.1142/S0218863595000070} {\bibfield  {journal} {\bibinfo  {journal} {J.
  Nonlinear Opt. Phys. Mater.}\ }\textbf {\bibinfo {volume} {4}},\ \bibinfo
  {pages} {141} (\bibinfo {year} {1995})}\BibitemShut {NoStop}%
\bibitem [{\citenamefont {Butov}\ and\ \citenamefont
  {Kavokin}(2012)}]{butov_behaviour_2012}%
  \BibitemOpen
  \bibfield  {author} {\bibinfo {author} {\bibfnamefont {L.~V.}\ \bibnamefont
  {Butov}}\ and\ \bibinfo {author} {\bibfnamefont {A.~V.}\ \bibnamefont
  {Kavokin}},\ }\href {\doibase 10.1038/nphoton.2011.325} {\bibfield  {journal}
  {\bibinfo  {journal} {Nat. Photon.}\ }\textbf {\bibinfo {volume} {6}},\
  \bibinfo {pages} {2} (\bibinfo {year} {2012})}\BibitemShut {NoStop}%
\bibitem [{\citenamefont
  {{Deveaud-Plédran}}(2012)}]{deveaud-pledran_condensation_2012}%
  \BibitemOpen
  \bibfield  {author} {\bibinfo {author} {\bibfnamefont {B.}~\bibnamefont
  {{Deveaud-Plédran}}},\ }\href {\doibase 10.1364/JOSAB.29.00A138} {\bibfield
  {journal} {\bibinfo  {journal} {J. Opt. Soc. Am. B}\ }\textbf {\bibinfo
  {volume} {29}},\ \bibinfo {pages} {A138} (\bibinfo {year}
  {2012})}\BibitemShut {NoStop}%
\bibitem [{\citenamefont {Klaers}\ \emph {et~al.}(2010)\citenamefont {Klaers},
  \citenamefont {Schmitt}, \citenamefont {Vewinger},\ and\ \citenamefont
  {Weitz}}]{klaers_boseeinstein_2010}%
  \BibitemOpen
  \bibfield  {author} {\bibinfo {author} {\bibfnamefont {J.}~\bibnamefont
  {Klaers}}, \bibinfo {author} {\bibfnamefont {J.}~\bibnamefont {Schmitt}},
  \bibinfo {author} {\bibfnamefont {F.}~\bibnamefont {Vewinger}}, \ and\
  \bibinfo {author} {\bibfnamefont {M.}~\bibnamefont {Weitz}},\ }\href
  {\doibase 10.1038/nature09567} {\bibfield  {journal} {\bibinfo  {journal}
  {Nature}\ }\textbf {\bibinfo {volume} {468}},\ \bibinfo {pages} {545–548}
  (\bibinfo {year} {2010})}\BibitemShut {NoStop}%
\bibitem [{\citenamefont {Bajoni}\ \emph {et~al.}(2007)\citenamefont {Bajoni},
  \citenamefont {Senellart}, \citenamefont {Lemaître},\ and\ \citenamefont
  {Bloch}}]{bajoni_photon_2007}%
  \BibitemOpen
  \bibfield  {author} {\bibinfo {author} {\bibfnamefont {D.}~\bibnamefont
  {Bajoni}}, \bibinfo {author} {\bibfnamefont {P.}~\bibnamefont {Senellart}},
  \bibinfo {author} {\bibfnamefont {A.}~\bibnamefont {Lemaître}}, \ and\
  \bibinfo {author} {\bibfnamefont {J.}~\bibnamefont {Bloch}},\ }\href
  {\doibase 10.1103/PhysRevB.76.201305} {\bibfield  {journal} {\bibinfo
  {journal} {Phys.~Rev.~B}\ }\textbf {\bibinfo {volume} {76}},\ \bibinfo
  {pages} {201305} (\bibinfo {year} {2007})}\BibitemShut {NoStop}%
\bibitem [{\citenamefont {Maragkou}\ \emph {et~al.}(2010)\citenamefont
  {Maragkou}, \citenamefont {Grundy}, \citenamefont {Ostatnický},\ and\
  \citenamefont {Lagoudakis}}]{maragkou_longitudinal_2010}%
  \BibitemOpen
  \bibfield  {author} {\bibinfo {author} {\bibfnamefont {M.}~\bibnamefont
  {Maragkou}}, \bibinfo {author} {\bibfnamefont {A.~J.~D.}\ \bibnamefont
  {Grundy}}, \bibinfo {author} {\bibfnamefont {T.}~\bibnamefont {Ostatnický}},
  \ and\ \bibinfo {author} {\bibfnamefont {P.~G.}\ \bibnamefont {Lagoudakis}},\
  }\href {\doibase 10.1063/1.3488012} {\bibfield  {journal} {\bibinfo
  {journal} {App.~Phys.~Lett.}\ }\textbf {\bibinfo {volume} {97}},\ \bibinfo
  {pages} {111110} (\bibinfo {year} {2010})}\BibitemShut {NoStop}%
\bibitem [{\citenamefont {Kammann}\ \emph {et~al.}(2011)\citenamefont
  {Kammann}, \citenamefont {Ohadi}, \citenamefont {Maragkou}, \citenamefont
  {Kavokin},\ and\ \citenamefont {Lagoudakis}}]{kammann_crossover_2011}%
  \BibitemOpen
  \bibfield  {author} {\bibinfo {author} {\bibfnamefont {E.}~\bibnamefont
  {Kammann}}, \bibinfo {author} {\bibfnamefont {H.}~\bibnamefont {Ohadi}},
  \bibinfo {author} {\bibfnamefont {M.}~\bibnamefont {Maragkou}}, \bibinfo
  {author} {\bibfnamefont {A.~V.}\ \bibnamefont {Kavokin}}, \ and\ \bibinfo
  {author} {\bibfnamefont {P.~G.}\ \bibnamefont {Lagoudakis}},\ }\href
  {http://arxiv.org/abs/1103.4831} {\bibfield  {journal} {\bibinfo  {journal}
  {{arXiv:1103.4831}}\ } (\bibinfo {year} {2011})}\BibitemShut {NoStop}%
\bibitem [{\citenamefont {Nelsen}\ \emph {et~al.}(2009)\citenamefont {Nelsen},
  \citenamefont {Balili}, \citenamefont {Snoke}, \citenamefont {Pfeiffer},\
  and\ \citenamefont {West}}]{nelsen_lasing_2009}%
  \BibitemOpen
  \bibfield  {author} {\bibinfo {author} {\bibfnamefont {B.}~\bibnamefont
  {Nelsen}}, \bibinfo {author} {\bibfnamefont {R.}~\bibnamefont {Balili}},
  \bibinfo {author} {\bibfnamefont {D.~W.}\ \bibnamefont {Snoke}}, \bibinfo
  {author} {\bibfnamefont {L.}~\bibnamefont {Pfeiffer}}, \ and\ \bibinfo
  {author} {\bibfnamefont {K.}~\bibnamefont {West}},\ }\href {\doibase
  doi:10.1063/1.3140822} {\bibfield  {journal} {\bibinfo  {journal}
  {J.~App.~Phys.}\ }\textbf {\bibinfo {volume} {105}},\ \bibinfo {pages}
  {122414} (\bibinfo {year} {2009})}\BibitemShut {NoStop}%
\bibitem [{\citenamefont {Tempel}\ \emph {et~al.}(2012)\citenamefont {Tempel},
  \citenamefont {Veit}, \citenamefont {Aßmann}, \citenamefont {Kreilkamp},
  \citenamefont {{Rahimi-Iman}}, \citenamefont {Löffler}, \citenamefont
  {Höfling}, \citenamefont {Reitzenstein}, \citenamefont {Worschech},
  \citenamefont {Forchel},\ and\ \citenamefont
  {Bayer}}]{tempel_characterization_2012}%
  \BibitemOpen
  \bibfield  {author} {\bibinfo {author} {\bibfnamefont {J.}~\bibnamefont
  {Tempel}}, \bibinfo {author} {\bibfnamefont {F.}~\bibnamefont {Veit}},
  \bibinfo {author} {\bibfnamefont {M.}~\bibnamefont {Aßmann}}, \bibinfo
  {author} {\bibfnamefont {L.~E.}\ \bibnamefont {Kreilkamp}}, \bibinfo {author}
  {\bibfnamefont {A.}~\bibnamefont {{Rahimi-Iman}}}, \bibinfo {author}
  {\bibfnamefont {A.}~\bibnamefont {Löffler}}, \bibinfo {author}
  {\bibfnamefont {S.}~\bibnamefont {Höfling}}, \bibinfo {author}
  {\bibfnamefont {S.}~\bibnamefont {Reitzenstein}}, \bibinfo {author}
  {\bibfnamefont {L.}~\bibnamefont {Worschech}}, \bibinfo {author}
  {\bibfnamefont {A.}~\bibnamefont {Forchel}}, \ and\ \bibinfo {author}
  {\bibfnamefont {M.}~\bibnamefont {Bayer}},\ }\href {\doibase
  10.1103/PhysRevB.85.075318} {\bibfield  {journal} {\bibinfo  {journal}
  {Phys.~Rev.~B}\ }\textbf {\bibinfo {volume} {85}},\ \bibinfo {pages} {075318}
  (\bibinfo {year} {2012})}\BibitemShut {NoStop}%
\bibitem [{\citenamefont {Lagoudakis}\ \emph {et~al.}(2002)\citenamefont
  {Lagoudakis}, \citenamefont {Savvidis}, \citenamefont {Baumberg},
  \citenamefont {Whittaker}, \citenamefont {Eastham}, \citenamefont
  {Skolnick},\ and\ \citenamefont {Roberts}}]{lagoudakis_stimulated_2002}%
  \BibitemOpen
  \bibfield  {author} {\bibinfo {author} {\bibfnamefont {P.~G.}\ \bibnamefont
  {Lagoudakis}}, \bibinfo {author} {\bibfnamefont {P.~G.}\ \bibnamefont
  {Savvidis}}, \bibinfo {author} {\bibfnamefont {J.~J.}\ \bibnamefont
  {Baumberg}}, \bibinfo {author} {\bibfnamefont {D.~M.}\ \bibnamefont
  {Whittaker}}, \bibinfo {author} {\bibfnamefont {P.~R.}\ \bibnamefont
  {Eastham}}, \bibinfo {author} {\bibfnamefont {M.~S.}\ \bibnamefont
  {Skolnick}}, \ and\ \bibinfo {author} {\bibfnamefont {J.~S.}\ \bibnamefont
  {Roberts}},\ }\href {\doibase 10.1103/PhysRevB.65.161310} {\bibfield
  {journal} {\bibinfo  {journal} {Phys.~Rev.~B}\ }\textbf {\bibinfo {volume}
  {65}},\ \bibinfo {pages} {161310} (\bibinfo {year} {2002})}\BibitemShut
  {NoStop}%
\bibitem [{\citenamefont {Kavokin}\ \emph {et~al.}(2003)\citenamefont
  {Kavokin}, \citenamefont {Lagoudakis}, \citenamefont {Malpuech},\ and\
  \citenamefont {Baumberg}}]{kavokin_polarization_2003}%
  \BibitemOpen
  \bibfield  {author} {\bibinfo {author} {\bibfnamefont {A.}~\bibnamefont
  {Kavokin}}, \bibinfo {author} {\bibfnamefont {P.~G.}\ \bibnamefont
  {Lagoudakis}}, \bibinfo {author} {\bibfnamefont {G.}~\bibnamefont
  {Malpuech}}, \ and\ \bibinfo {author} {\bibfnamefont {J.~J.}\ \bibnamefont
  {Baumberg}},\ }\href {\doibase 10.1103/PhysRevB.67.195321} {\bibfield
  {journal} {\bibinfo  {journal} {Phys.~Rev.~B}\ }\textbf {\bibinfo {volume}
  {67}},\ \bibinfo {pages} {195321} (\bibinfo {year} {2003})}\BibitemShut
  {NoStop}%
\bibitem [{\citenamefont {Laussy}\ \emph {et~al.}(2006)\citenamefont {Laussy},
  \citenamefont {Shelykh}, \citenamefont {Malpuech},\ and\ \citenamefont
  {Kavokin}}]{laussy_effects_2006}%
  \BibitemOpen
  \bibfield  {author} {\bibinfo {author} {\bibfnamefont {F.~P.}\ \bibnamefont
  {Laussy}}, \bibinfo {author} {\bibfnamefont {I.~A.}\ \bibnamefont {Shelykh}},
  \bibinfo {author} {\bibfnamefont {G.}~\bibnamefont {Malpuech}}, \ and\
  \bibinfo {author} {\bibfnamefont {A.}~\bibnamefont {Kavokin}},\ }\href
  {\doibase 10.1103/PhysRevB.73.035315} {\bibfield  {journal} {\bibinfo
  {journal} {Phys.~Rev.~B}\ }\textbf {\bibinfo {volume} {73}},\ \bibinfo
  {pages} {035315} (\bibinfo {year} {2006})}\BibitemShut {NoStop}%
\bibitem [{\citenamefont {Read}\ \emph {et~al.}(2009)\citenamefont {Read},
  \citenamefont {Liew}, \citenamefont {Rubo},\ and\ \citenamefont
  {Kavokin}}]{read_stochastic_2009}%
  \BibitemOpen
  \bibfield  {author} {\bibinfo {author} {\bibfnamefont {D.}~\bibnamefont
  {Read}}, \bibinfo {author} {\bibfnamefont {T.~C.~H.}\ \bibnamefont {Liew}},
  \bibinfo {author} {\bibfnamefont {Y.~G.}\ \bibnamefont {Rubo}}, \ and\
  \bibinfo {author} {\bibfnamefont {A.~V.}\ \bibnamefont {Kavokin}},\ }\href
  {\doibase 10.1103/PhysRevB.80.195309} {\bibfield  {journal} {\bibinfo
  {journal} {Phys.~Rev.~B}\ }\textbf {\bibinfo {volume} {80}},\ \bibinfo
  {pages} {195309} (\bibinfo {year} {2009})}\BibitemShut {NoStop}%
\bibitem [{\citenamefont {Kasprzak}\ \emph {et~al.}(2007)\citenamefont
  {Kasprzak}, \citenamefont {André}, \citenamefont {Dang}, \citenamefont
  {Shelykh}, \citenamefont {Kavokin}, \citenamefont {Rubo}, \citenamefont
  {Kavokin},\ and\ \citenamefont {Malpuech}}]{kasprzak_build_2007}%
  \BibitemOpen
  \bibfield  {author} {\bibinfo {author} {\bibfnamefont {J.}~\bibnamefont
  {Kasprzak}}, \bibinfo {author} {\bibfnamefont {R.}~\bibnamefont {André}},
  \bibinfo {author} {\bibfnamefont {L.~S.}\ \bibnamefont {Dang}}, \bibinfo
  {author} {\bibfnamefont {I.~A.}\ \bibnamefont {Shelykh}}, \bibinfo {author}
  {\bibfnamefont {A.~V.}\ \bibnamefont {Kavokin}}, \bibinfo {author}
  {\bibfnamefont {Y.~G.}\ \bibnamefont {Rubo}}, \bibinfo {author}
  {\bibfnamefont {K.~V.}\ \bibnamefont {Kavokin}}, \ and\ \bibinfo {author}
  {\bibfnamefont {G.}~\bibnamefont {Malpuech}},\ }\href {\doibase
  10.1103/PhysRevB.75.045326} {\bibfield  {journal} {\bibinfo  {journal}
  {Phys.~Rev.~B}\ }\textbf {\bibinfo {volume} {75}},\ \bibinfo {pages} {045326}
  (\bibinfo {year} {2007})}\BibitemShut {NoStop}%
\bibitem [{\citenamefont {Christmann}\ \emph {et~al.}(2008)\citenamefont
  {Christmann}, \citenamefont {Butté}, \citenamefont {Feltin}, \citenamefont
  {Carlin},\ and\ \citenamefont {Grandjean}}]{christmann_room_2008}%
  \BibitemOpen
  \bibfield  {author} {\bibinfo {author} {\bibfnamefont {G.}~\bibnamefont
  {Christmann}}, \bibinfo {author} {\bibfnamefont {R.}~\bibnamefont {Butté}},
  \bibinfo {author} {\bibfnamefont {E.}~\bibnamefont {Feltin}}, \bibinfo
  {author} {\bibfnamefont {J.}~\bibnamefont {Carlin}}, \ and\ \bibinfo {author}
  {\bibfnamefont {N.}~\bibnamefont {Grandjean}},\ }\href {\doibase
  doi:10.1063/1.2966369} {\bibfield  {journal} {\bibinfo  {journal}
  {App.~Phys.~Lett.}\ }\textbf {\bibinfo {volume} {93}},\ \bibinfo {pages}
  {051102} (\bibinfo {year} {2008})}\BibitemShut {NoStop}%
\bibitem [{\citenamefont {Kappei}\ \emph {et~al.}(2005)\citenamefont {Kappei},
  \citenamefont {Szczytko}, \citenamefont {{Morier-Genoud}},\ and\
  \citenamefont {Deveaud}}]{kappei_direct_2005}%
  \BibitemOpen
  \bibfield  {author} {\bibinfo {author} {\bibfnamefont {L.}~\bibnamefont
  {Kappei}}, \bibinfo {author} {\bibfnamefont {J.}~\bibnamefont {Szczytko}},
  \bibinfo {author} {\bibfnamefont {F.}~\bibnamefont {{Morier-Genoud}}}, \ and\
  \bibinfo {author} {\bibfnamefont {B.}~\bibnamefont {Deveaud}},\ }\href
  {\doibase 10.1103/PhysRevLett.94.147403} {\bibfield  {journal} {\bibinfo
  {journal} {Phys.~Rev.~Lett.}\ }\textbf {\bibinfo {volume} {94}},\ \bibinfo
  {pages} {147403} (\bibinfo {year} {2005})}\BibitemShut {NoStop}%
\bibitem [{\citenamefont {Stern}\ \emph {et~al.}(2008)\citenamefont {Stern},
  \citenamefont {Garmider}, \citenamefont {Umansky},\ and\ \citenamefont
  {{Bar-Joseph}}}]{stern_mott_2008}%
  \BibitemOpen
  \bibfield  {author} {\bibinfo {author} {\bibfnamefont {M.}~\bibnamefont
  {Stern}}, \bibinfo {author} {\bibfnamefont {V.}~\bibnamefont {Garmider}},
  \bibinfo {author} {\bibfnamefont {V.}~\bibnamefont {Umansky}}, \ and\
  \bibinfo {author} {\bibfnamefont {I.}~\bibnamefont {{Bar-Joseph}}},\ }\href
  {\doibase 10.1103/PhysRevLett.100.256402} {\bibfield  {journal} {\bibinfo
  {journal} {Phys.~Rev.~Lett.}\ }\textbf {\bibinfo {volume} {100}},\ \bibinfo
  {pages} {256402} (\bibinfo {year} {2008})}\BibitemShut {NoStop}%
\bibitem [{\citenamefont {Shelykh}\ \emph {et~al.}(2004)\citenamefont
  {Shelykh}, \citenamefont {Kavokin}, \citenamefont {Kavokin}, \citenamefont
  {Malpuech}, \citenamefont {Bigenwald}, \citenamefont {Deng}, \citenamefont
  {Weihs},\ and\ \citenamefont {Yamamoto}}]{shelykh_semiconductor_2004}%
  \BibitemOpen
  \bibfield  {author} {\bibinfo {author} {\bibfnamefont {I.}~\bibnamefont
  {Shelykh}}, \bibinfo {author} {\bibfnamefont {K.~V.}\ \bibnamefont
  {Kavokin}}, \bibinfo {author} {\bibfnamefont {A.~V.}\ \bibnamefont
  {Kavokin}}, \bibinfo {author} {\bibfnamefont {G.}~\bibnamefont {Malpuech}},
  \bibinfo {author} {\bibfnamefont {P.}~\bibnamefont {Bigenwald}}, \bibinfo
  {author} {\bibfnamefont {H.}~\bibnamefont {Deng}}, \bibinfo {author}
  {\bibfnamefont {G.}~\bibnamefont {Weihs}}, \ and\ \bibinfo {author}
  {\bibfnamefont {Y.}~\bibnamefont {Yamamoto}},\ }\href {\doibase
  10.1103/PhysRevB.70.035320} {\bibfield  {journal} {\bibinfo  {journal}
  {Phys.~Rev.~B}\ }\textbf {\bibinfo {volume} {70}},\ \bibinfo {pages} {035320}
  (\bibinfo {year} {2004})}\BibitemShut {NoStop}%
\bibitem [{\citenamefont {Nardin}\ \emph {et~al.}(2009)\citenamefont {Nardin},
  \citenamefont {Lagoudakis}, \citenamefont {Wouters}, \citenamefont {Richard},
  \citenamefont {Baas}, \citenamefont {André}, \citenamefont {Dang},
  \citenamefont {Pietka},\ and\ \citenamefont
  {{Deveaud-Plédran}}}]{nardin_dynamics_2009}%
  \BibitemOpen
  \bibfield  {author} {\bibinfo {author} {\bibfnamefont {G.}~\bibnamefont
  {Nardin}}, \bibinfo {author} {\bibfnamefont {K.~G.}\ \bibnamefont
  {Lagoudakis}}, \bibinfo {author} {\bibfnamefont {M.}~\bibnamefont {Wouters}},
  \bibinfo {author} {\bibfnamefont {M.}~\bibnamefont {Richard}}, \bibinfo
  {author} {\bibfnamefont {A.}~\bibnamefont {Baas}}, \bibinfo {author}
  {\bibfnamefont {R.}~\bibnamefont {André}}, \bibinfo {author} {\bibfnamefont
  {L.~S.}\ \bibnamefont {Dang}}, \bibinfo {author} {\bibfnamefont
  {B.}~\bibnamefont {Pietka}}, \ and\ \bibinfo {author} {\bibfnamefont
  {B.}~\bibnamefont {{Deveaud-Plédran}}},\ }\href {\doibase
  10.1103/PhysRevLett.103.256402} {\bibfield  {journal} {\bibinfo  {journal}
  {Phys.~Rev.~Lett.}\ }\textbf {\bibinfo {volume} {103}},\ \bibinfo {pages}
  {256402} (\bibinfo {year} {2009})}\BibitemShut {NoStop}%
\end{thebibliography}%

\end{document}